# The role of fingerprints in the coding of tactile information probed with a biomimetic sensor

J. Scheibert,[1] S. Leurent,[1] A. Prevost,[1] G. Debrégeas[1,2]

**In humans, the tactile perception of fine textures (spatial scale <200µm) is mediated by skin vibrations generated as the finger scans the surface. To establish the relationship between texture characteristics and subcutaneous vibrations, a biomimetic tactile sensor has been designed whose dimensions match those of the fingertip. When the sensor surface is patterned with parallel ridges mimicking the fingerprints, the spectrum of vibrations elicited by randomly textured substrates is dominated by one frequency set by the ratio of the scanning speed to the inter-ridge distance. For human touch, this frequency falls within the optimal range of sensitivity of Pacinian afferents which mediate the coding of fine textures. Thus, fingerprints may perform spectral selection and amplification of tactile information which facilitate its processing by specific mechanoreceptors.**

[1] Laboratoire de Physique Statistique de l'ENS, UMR 8550, CNRS-ENS-Université Paris 6 & Paris 7, 24 rue Lhomond F-75231 Paris, France.
[2] To whom correspondence should be addressed; Email: georges.debregeas@lps.ens.fr.

The hand is an important means for human interaction with the physical environment (*1*). Many of the tasks that the hand can undertake - such as precision grasping and manipulation of objects, detection of individual defects on smooth surfaces, discrimination of textures, etc. - depend on the exquisite tactile sensitivity of the fingertips. Tactile information is conveyed by populations of mechanosensitive afferent fibers innervating the distal fingerpads (*2, 3*). In recent years, a breakthrough in our understanding of the coding of roughness perception has been made with the experimental confirmation of Katz' historical proposition of the existence of two independent coding channels that are specific for the perception of coarse and fine textures (*4-6*). The perception of coarse textures (with features of lateral dimensions larger than about $200 \mu m$) relies on spatial variations of the finger/substrate contact stress field and is mediated by the slowly adapting mechanoreceptors (*7*). In contrast, the perception of finer textures ($< 200 \mu m$) requires the finger to be scanned across the surface since it is based on the cutaneous vibrations thus elicited. These vibrations are intensively encoded principally by Pacinian fibers (*8*) which are characterized by a band-pass behavior with a best frequency (i.e. the stimulus frequency where maximum sensitivity occurs) of order $250 Hz$ (*9*). The most elaborated description of the latter coding scheme was given by Bensmaïa and Hollins who directly measured the skin vibrations of fingers scanning finely textured substrates. They were able to correlate the perceived roughness of the surface with the power of the texture-induced vibrations weighted by the Pacinian spectral sensitivity (*10, 11*).

Among the four types of mechanoreceptors that convey tactile information, Pacinian corpuscles (PC's) have the most extended receptive field and therefore the lowest spatial resolution. This may seem paradoxical given their involvement in the tactile perception of fine features (*12, 13*). In standard psychophysical tests, the substrates used as stimuli are made of regularly spaced dots or bars (*1*). The resulting skin vibrations are thus confined to a single frequency whose value can be actively tuned by the subject through the scanning velocity so that it falls within the PC optimal range of sensitivity. Such regular stimuli substrates thus favor tactile identification or classification tasks.   In contrast, for natural surfaces where features are randomly distributed and exhibit a wide spectrum of size, the elicited skin vibrations are expected to be spread over a large range of frequencies among which only a limited fraction contributes to the PC activity.

To address this question on how low-resolution receptors encode fine textural information, the present study investigates the mechanical filtering properties of the skin. It aims at characterizing how textural information at any spatial scale (less than the finger/substrate contact diameter) is converted into subcutaneous vibrations in the vicinity of the mechanoreceptors during a dynamic tactile exploration. Since there is currently no way to measure experimentally the subcutaneous stress using a human subject, our approach is based on the use of a biomimetic tactile sensor whose functioning principle and main geometrical characteristics are matched to those of the human fingertip. This allows us in particular to test the role of epidermal ridges (fingerprints) in this transduction process. Two distinct functional roles have been so far attributed to these characteristic structures of the digital skin. Fingerprints are believed to reinforce friction and adhesion of the fingerpads thus improving the ability to securely grasp objects or supports (*14*, *1*). They may also be implicated in tactile perception, each of them acting as a magnifying lever thus increasing the subsurface strain with respect to the surface deformation (*15*, *16*). Here we show that fingerprints may have a strong impact on the spectral filtering properties of the skin in dynamic tactile exploration.

The tactile sensor aims at mimicking the operation of the PC in dynamic tactile exploration (*17*, *18*). As far as possible, the various geometrical and mechanical characteristics of the sensor are scaled to its biological counterpart (see Fig. S1 for a comparison of key parameters). The sensing element consists of a MEMS (Micro-Electro Mechanical System) device which provides force measurements in a region of millimeter extension. This micro-force sensor is attached to a rigid base and covered with an elastic spherical cap mimicking the fingertip skin (Fig. 1A). This cap, made of cross-linked PDMS (Poly[DimethylSiloxane]), has a maximum thickness $h = 2\,\text{mm}$. Its surface is either "smooth" or "fingerprinted", i.e. patterned with a regular square wave grating of period $\lambda = 220\,\mu\text{m}$ and depth $28\,\mu\text{m}$. The tactile sensor is mounted on a double cantilever system allowing one to record the normal and tangential loads using capacitive position sensors. In a typical experiment, the sensor is scanned at constant velocity across a rigid, nominally flat substrate under a constant normal load $P = 1.71\,\text{N}$ yielding a contact zone of centimeter extension. This value for the load, together with the periodicity of the fingerprint-like structure, is chosen so that the number of ridges within the contact in the artificial system is close to that observed with an actual fingerpad under standard exploratory load (as illustrated in Fig. 1B and 1C).

The stimuli consist of white-noise 1D textured substrates (Fig. 1A-upper inset). They are obtained by patterning glass slides with a $28\,\mu\text{m}$ deep square wave grating whose edges are positioned at random positions with a mean grating width of $75\,\mu\text{m}$ (17). The fingerprint-like ridges (when present) and substrate gratings are parallel to each other and oriented perpendicularly to the sliding direction. For moderate scanning velocities ($v < 0.4\,\text{mm/s}$) and a given normal load, the pressure signal $p(t)$ is found to be a sole function of the substrate position at time $t$ regardless of the scanning velocity $v$ (Fig. S2 and S3). All experiments are performed at constant $v = 0.2\,\text{mm/s}$ well within this velocity-independent regime of friction. To facilitate the analysis, data are systematically plotted as a function of the sensor/substrate relative displacement $u = v\,t$ since a strict equivalence exists between time and substrate displacement in steady sliding.

Figure 2A shows the typical pressure variations $p(u) - \langle p \rangle$ (where $\langle p \rangle$ is the average pressure) measured with the micro-force device as the sensor is scanned across a textured surface. The smooth sensor exhibits pressure modulations with a characteristic wavelength in the millimeter range. The fingerprinted system reveals similar long-wavelength modulations to which are superimposed rapid oscillations whose period corresponds to a displacement of the substrate over the inter-ridge distance $\lambda = 220\mu\mathrm{m}$. A characterization of both sensors' filtering properties is given by Fig. 2B which displays the power spectra of both signals together with that of the input stimulus, i.e. the substrate topography (dashed line). The smooth sensor acts as a low-pass filter since it rapidly attenuates all pressure modulations induced by texture components of wavelength smaller than $\approx 1\mathrm{mm}$. In contrast, the fingerprinted sensor exhibits band-pass filtering characteristics around the spatial frequency $1/\lambda$ (with further harmonics at integer multiples of $1/\lambda$). The presence of fingerprint-like ridges results in an amplification by a factor 100 of the pressure modulations induced by a texture of wavelength $\lambda$ (19).

These filtering characteristics can be interpreted to first order using a linear mechanical description of tactile sensing (20). Consider a small linear force sensor embedded at a depth $h$ in an elastic skin and located at $(x = 0, y = 0)$. Its response to localized unit forces applied at various positions $(x, y)$ on the skin surface defines its receptive field $F(x, y)$. The sensor signal $p$ induced by any stress field $\sigma^s(x, y)$ applied at the skin surface then reads $p = \iint F(x, y)\sigma^s(x, y)\, dx\, dy$. We denote $\bar{\sigma}(x, y)$ the (time invariant) contact stress field resulting from the continuous rubbing of a smooth substrate under a given load. If the substrate exhibits a fine texture, the stress field $\sigma^s$ becomes dependent on the substrate position $u$. As $u$ varies, $\sigma^s$ is modulated around the reference field $\bar{\sigma}(x, y)$. The use of substrates exhibiting a two-level topography and a large enough contrast prevents from any contact above the wells (as optically evidenced in Fig. S4). The contact pressure is thus zero over half of the apparent contact region whereas it is expected to be of order twice the time-averaged stress field $\overline{\sigma}(x, y)$ at the location of the substrate summits. As a first approximation, one may thus write the superficial stress field as a function of $u$ in the form

$$\sigma^s(x, y) = \overline{\sigma}(x, y).(1 + T(u - x))$$ (1)

where $T(x)$ is the normalized binary function ($T = \pm 1$) representing the topography of the surface. It should be noted that an exact calculation of the contact stress at a given location should take into account the local topography of the substrate and not only the average fraction of summits. The induced corrections should be significant at short length-scales but become small when considering stress modulations over distances larger than the mean grating period.

With this expression, the pressure signal is then given by

$$p(u) = \langle p \rangle + \iint (F.\overline{\sigma})(x, y).T(u - x) dx dy$$ (2)

The transduction of tactile information is controlled by the product of the receptive field $F$ and the reference stress field $\overline{\sigma}$. The function $F$ characterizes the intrinsic properties of the receptor. It is expected to have a typical lateral extension of order $h$ and to be fairly independent of the skin topography (such as fingerprints) provided that the height of the surface features is less than $h$ (*21*). The reference field $\overline{\sigma}$ depends on the exploratory conditions such as the normal load $P$, the friction coefficient or the position of the contact zone with respect to the sensor location. Unlike $F$, the stress field $\overline{\sigma}$ is highly sensitive to the skin surface topography. In particular, the presence of fingerprints a few tens of micrometers deep leads to a complete extinction of $\overline{\sigma}$ along regularly spaced lines (as illustrated in Fig. S6), resulting in the observed spectral amplification of the signal at the frequency $1/\lambda$.

Equation 2 can be re-written as $p(u) = \langle p \rangle + \int g_1(x)T(u-x)dx$ where $g_1(x) = \int (F.\overline{\sigma})(x,y)dy$ now defines the linear response function of the sensor with respect to 1D two-levels stimuli substrates. The use of white-noise stimuli enables us to implement a Wiener-Volterra reverse-correlation method and extract $g_1(x)$ directly from the measurements, $g_1(x) = \langle p(u)T(u-x) \rangle$ (*22*, *23*). The result of this computation for both smooth and fingerprinted sensors is plotted on Fig. 3. In qualitative agreement with the linear model, both response functions exhibit an envelope of lateral extension of order $h$ and the response function of the fingerprinted sensor is further modulated with a spatial period $\lambda$. These functions can be tested by confronting actual measurements of $p(u) - \langle p \rangle$ with the predicted signal $\int g_1(x)T(u-x)dx$ as shown in Fig. 4A for the fingerprinted system. To facilitate the comparison, Fig. 4B and 4C display the low- and high-frequency components, respectively. The linear response function allows one to reproduce the low-frequency signal. Although it correctly predicts the maxima and minima of the high-frequency component, it fails to capture its amplitude which indicates that non-linear effects might not be negligible for small length-scales. These effects could be taken into account by correlating $p$ with the successive powers of $T$ in order to include additional terms of the Wiener-Volterra series to describe the response function. However, this computation would require using a much larger set of stimuli to provide sufficient statistics.

Although the biomimetic tactile sensor used in this study offers a crude version of the finger physiology (*24*, *25*), the mechanism of spectral selection it helped unravel depends on a very limited set of ingredients and should therefore be relevant to human digital touch. Namely, it requires that the surface of the tactile sensor displays a regularly ridged topography whose spatial period and amplitude are much smaller than the receptive field diameter and the mechanoreceptor's depth. In these conditions, such ridges have little influence on the skin deformations induced by a coarse texture (of spatial scale larger than the inter-ridge distance $\lambda$). However, by shaping the interfacial contact stress field, such epidermal ridges give rise to an amplification of the subsurface stress modulations induced by a texture of characteristic wavelength equal to $\lambda$. In the time domain, this spatial period corresponds to a frequency $f_0 = v / \lambda$ where $v$ is the finger/substrate relative velocity. In natural exploratory conditions, $v$ is observed to be of order $10 - 15\,\text{cm/s}$ (*1*). With a typical inter-ridge distance $\lambda \approx 500\,\mu\text{m}$, this yields a frequency $f_0 \approx 200 - 300\,\text{Hz}$ of the order of the best frequency of the Pacinian fibers which mediate the coding of fine textures. Fingerprints thus allow for a conditioning of the texture-induced mechanical signal which facilitates its processing by specific mechanoreceptors. It should be noted that this process is strongly dependent on the orientation of the ridges with respect to the scanning direction (Fig. S7). In humans, fingerprints are organized in elliptical twirls so that each region of the fingertip (and thus each PC) can be ascribed with an optimal scanning orientation. Further studies are needed in order to elucidate how this may reflect on the exploratory procedures (such as fingertip trajectory and contacting zone) used by humans during texture evaluation tasks.

Remarkably, the response function of the fingerprinted system displayed in Fig. 3 is analogous to a Gabor filter since it provides both spatial and spectral resolution. Such filters are classically used in image analysis and have been identified in visual systems at the neural level (*26*). They are known to provide orientation discrimination, contrast enhancement and motion detection. One may therefore expects, beyond the spectral filtering process discussed here, other interesting functional consequences of fingerprints, presumably relevant to the design of realistic haptic interfaces for humanoid robots (*27*, *28*).

29. This project was supported mostly by CNRS basic funding and in part by the EU-NEST programme, MONAT project (contract 21 number 29000). We are grateful to Didier Chatenay and Laurent Bourdieu for fruitful discussions and careful reading of the manuscript.

# Figures

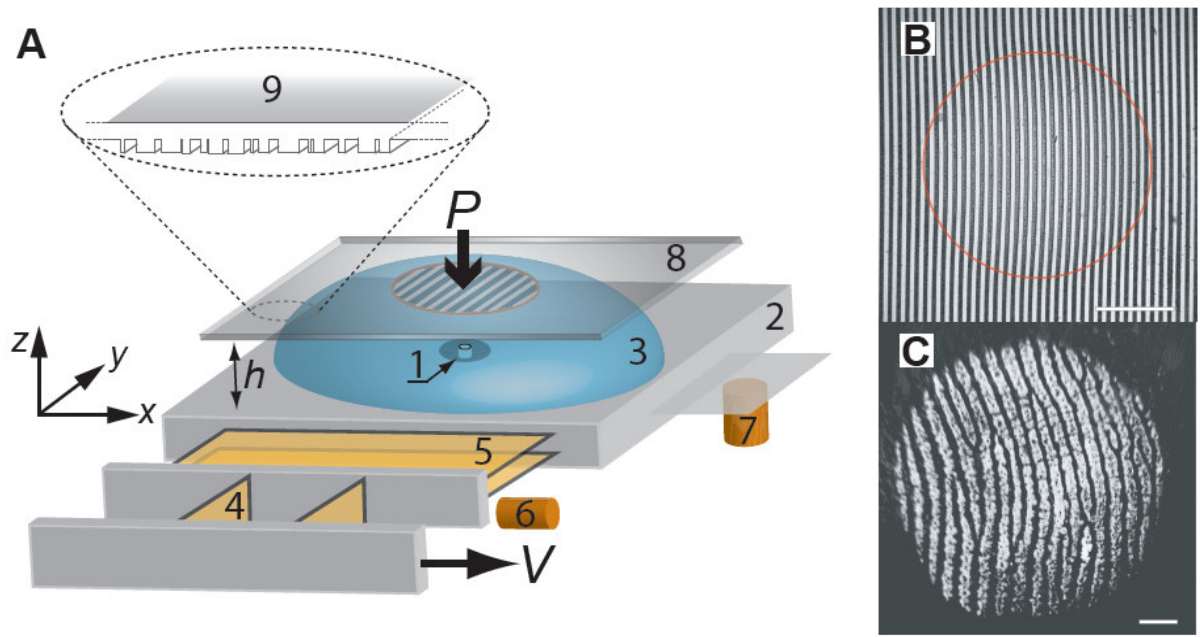

**Fig. 1. (A)** Sketch of the experimental setup. A MEMS micro-force sensor (1) is mounted on a rigid base (2). It is covered with a spherical elastomer cap (3) of maximum thickness $h = 2\,\text{mm}$ and whose surface is smooth or patterned with parallel ridges. The resulting tactile sensor is mounted on a double cantilever system (4, 5) allowing one to measure the total normal and tangential loads exerted on the sensor using capacitive position sensors (6, 7). In a typical experiment, the tactile sensor is scanned at constant speed $v$ (using a linear motor) and under constant normal load $P$, across glass slides (8) whose surface is patterned with a 1D random square-wave grating (9). **(B)** Snapshot of the contact between the fingerprinted cap and a smooth glass slide in steady sliding. Wells between the elastomer's ridges appear bright and the red solid line circle, also shown on (A), defines the border of the contact. Actual contact only occurs on the ridges summits. Ridges are slightly deformed around the contact due to interfacial friction. **(C)** For comparison, this snapshot displays the contact between a human fingertip and a smooth glass surface with $P \approx 0.5\,\text{N}$ (a typical value in tactile exploration). In both (B) and (C), the white bar is $2\,\text{mm}$ long.

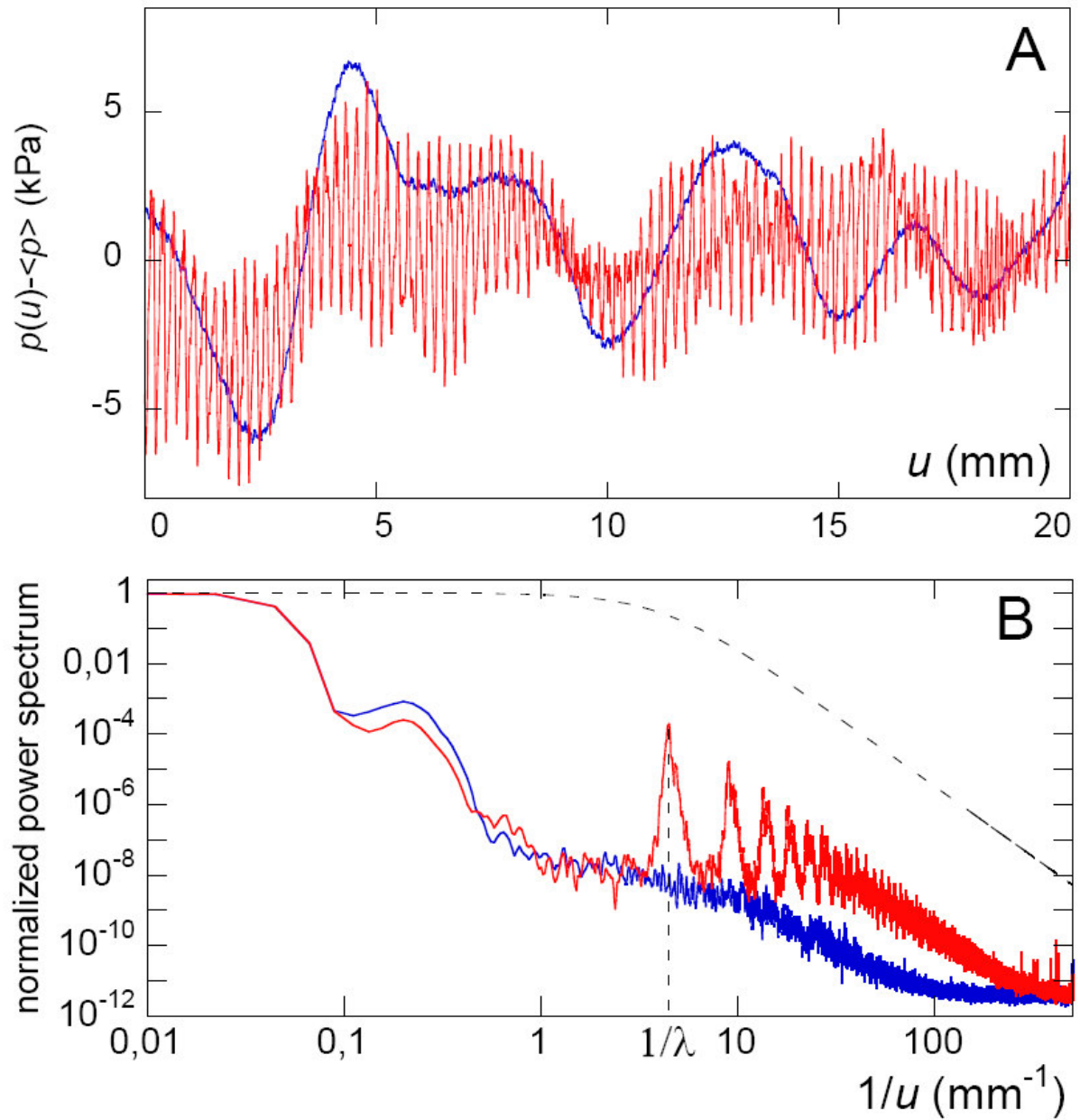

**Fig. 2.** **(A)** Typical pressure variation $p - \langle p \rangle$ measured with the smooth (blue) and fingerprinted (red) biomimetic fingers as a function of the substrate displacement $u$. The stimulus substrate used to produce these signals is a patterned glass slide exhibiting 1D random roughness. **(B)** Normalized power spectra of both signals obtained by Fourier transform averaged over 4 data sets, equivalent to a substrate of total length 180mm. Shown in dashed lines is the theoretical power spectrum of the random pattern used as stimuli.

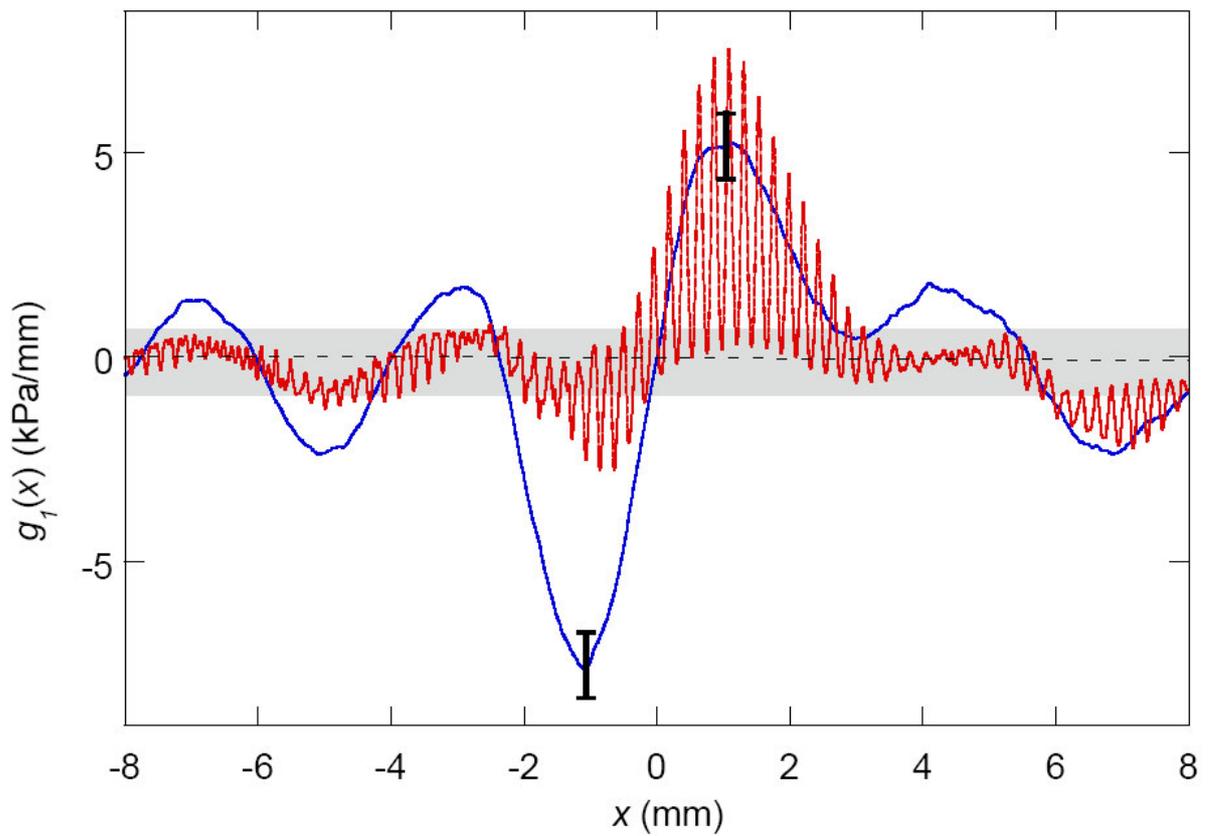

**Fig. 3.** Linearized stimulus-signal response functions $g_1(x)$ computed by cross-correlating the pressure signals and the stimulus topography $T(x)$, for both smooth (blue) and fingerprinted (red) systems. These data were obtained by averaging over 3 data sets, each one corresponding to a substrate length of 45mm. The expected statistical deviation due to the finite length of the substrates was estimated numerically to be $\pm 0.75 \text{kPa/mm}$. This value is shown with the error bars and the shaded rectangle.

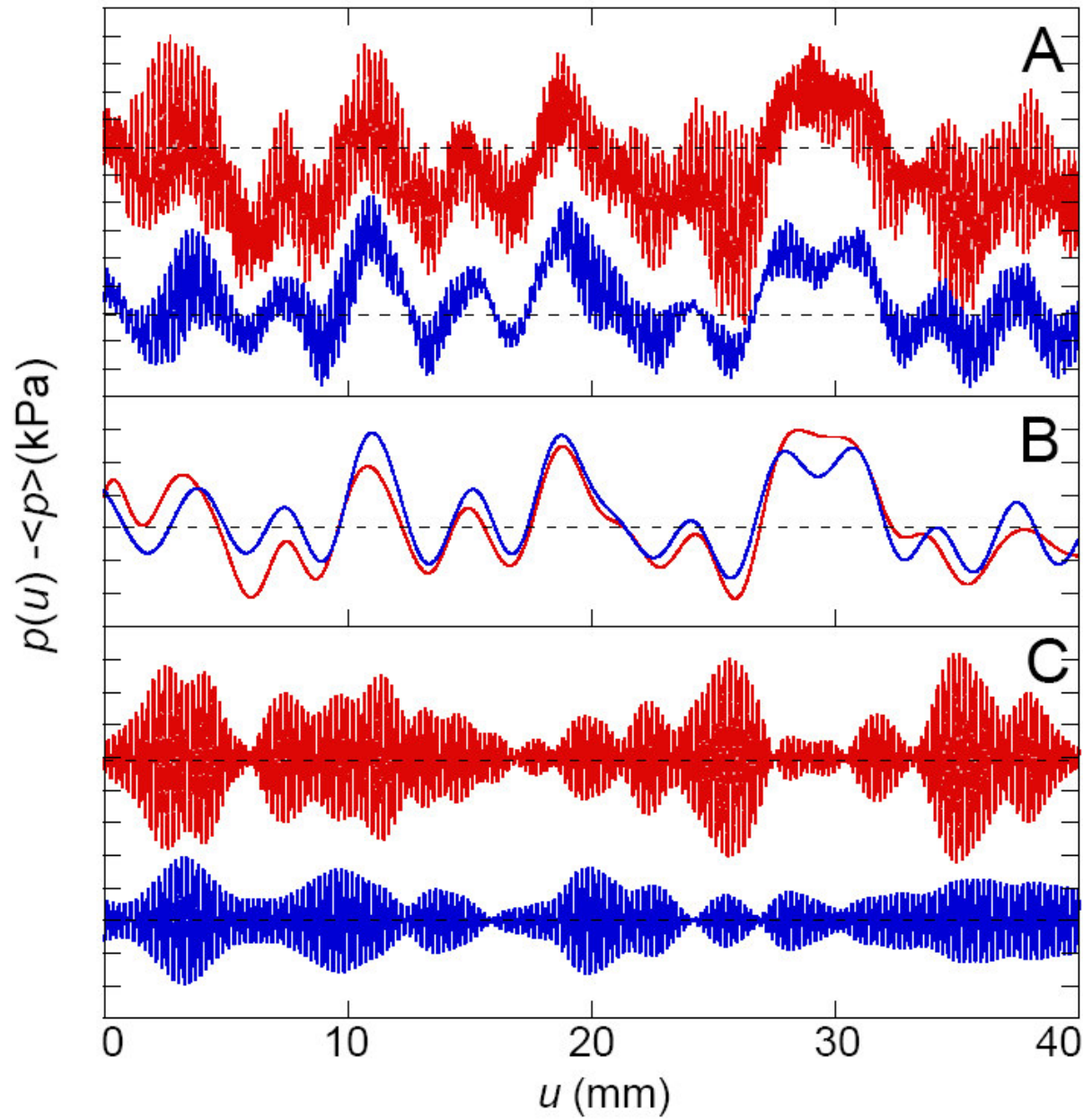

**Fig. 4.** **(A)** In red, pressure signal $p - \langle p \rangle$ measured with the fingerprinted sensor on a rough substrate. In blue, predicted signal obtained by convoluting the substrate topography function $T(x)$ with the linear response function $g_1(x)$. The latter was obtained independently by reverse-correlation using 2 distinct $45\,\mathrm{mm}$-long substrates. The dotted line indicates the $y = 0$ axis and each interval along the $y$-axis corresponds to a pressure variation of $1\,\mathrm{kPa}$. For easier comparison, the same signals are plotted after applying **(B)** a low-pass filter with a cutoff frequency of $1/(2\lambda)$ and **(C)** a band-pass filter centered around the peak frequency $1/\lambda$.

# Supporting Material

# The role of fingerprints in the coding of tactile information probed with a biomimetic sensor

J. Scheibert, S. Leurent, A. Prevost, G. Debrégeas

# I – Materials and methods

## 1- Design of the biomimetic sensors

The principle and calibration of the biomimetic sensor have been described in a previous publication (*S1*). The sensing element is a MEMS (Micro-Electro Mechanical System) device designed by LETI (CEA, Grenoble, France). It allows for the measurements of the three components of the local force in a region of millimetric extension. In this article, only the normal component (local pressure $p$) was analysed. "Smooth" and "fingerprinted" membranes were made out of an optically transparent PDMS elastomer (Poly[DimethylSiloxane], Sylgard 184, Dow Corning) of elastic modulus $2.2 \pm 0.1$ MPa. Their spherical shape was obtained by filling, prior to curing, a plano-concave spherical glass lens (radius of curvature 128.8 mm) with the liquid PDMS-crosslinker mixture.

To reduce the adhesion and friction coefficients of the membranes against the substrates (and avoid damages to the micro-force sensor) the concave lens surface was finely abraded with a liquid water-SiC powder which, after molding, resulted in a mat finish of the elastomer surface. To limit residual stress, curing was performed at room temperature for at least 48 hours, after which the elastomer cap was peeled off from its cast and "glued" on top of the micro-sensor using a thin PDMS-crosslinker liquid film. The "fingerprinted" membrane was designed by soft photolithography. A layer of photoresist (SU8-2035, Microchem Inc) was spin-coated on the abraded lens, and UV exposed through a mask consisting of alternating opaque and transparent parallel stripes of equal width 110µm. After development, one was left with a grating pattern of parallel grooves 28µm deep with the ridges summits displaying a mat finish similar to the smooth membrane.

The tactile sensor main characteristics (membrane dimensions and rigidity, micro-force sensor's sensitive area) are comparable to the physiological system ones as shown in Fig. S1.

## 2- Fabrication of the rough substrates

The substrates used as stimuli consisted of 1D square wave gratings designed with similar lithography techniques as detailed above. They were produced by patterning 28µm thick layers of SU8-2035 photoresist spin coated on microscope glass slides ($26 \times 76$ mm). The masks were designed with a bar code like pattern consisting of successive and alternating opaque and transparent stripes, 70mm long, whose edges locations were chosen from a uniform distribution (Fig. S4). This procedure resulted in a low pass white noise power spectrum with a cut-off spatial frequency $1/(\pi l)$ where $l$ = 75µm is the mean distance between successive edges. The profile $T(x)$ of the surface topography was extracted by optical profilometry (M3D, Fogale Nanotech).

As a test of robustness of the observed effect, a series of experiments was run using abraded substrates (Fig.S5) obtained by mechanical roughening of microscope glass slides with a liquid water-SiC powder (mean particle diameter 37µm). The surface topography displayed root mean square (rms) roughness of 1.2µm as measured by optical profilometry.

## 3- Friction experiments

Experiments were carried out using a frictional setup described in (*S1*). The bio-mimetic finger was mounted on a double cantilever system allowing one to record the normal and tangential loads using capacitive position sensors (Fogale Nanotech). The set-up was driven at a constant speed using a DC linear motor (LTA-HS actuator, Newport

Inc.). Precision translation and tilt stages allowed for micrometric positioning of the sensor with respect to the substrates. Force measurements were recorded onto a hard drive using an A/D board (PCI 6255, 16 bits, National Instruments) and later analyzed.

In a typical experiment, the tactile sensor was first pressed against the substrate up to the prescribed normal load. Force signals were recorded as the tactile sensor was moved for 50 mm along the substrate under constant normal force and scanning velocity. Misalignment between the substrate and the axis of motion generally resulted in a drift of the measured normal force. Once corrected, the normal force was found to vary by less than 1% over the whole substrate. Ten experiments were then carried out over the same region of the substrate to guarantee the reproducibility of the data set (Fig. S2).

In all experiments described in the manuscript, the ridges are parallel to the substrate grating and perpendicular to the scanning direction. To probe the effect of such an alignment on the amplitude of the measured signal, a series of experiments has been carried out where the orientation of the ridges was gradually tilted with respect to the scanning direction and substrate grating axis. This was achieved by mounting the tactile sensor on a rotating stage (Fig. S7).

# II – Supplementary Figures

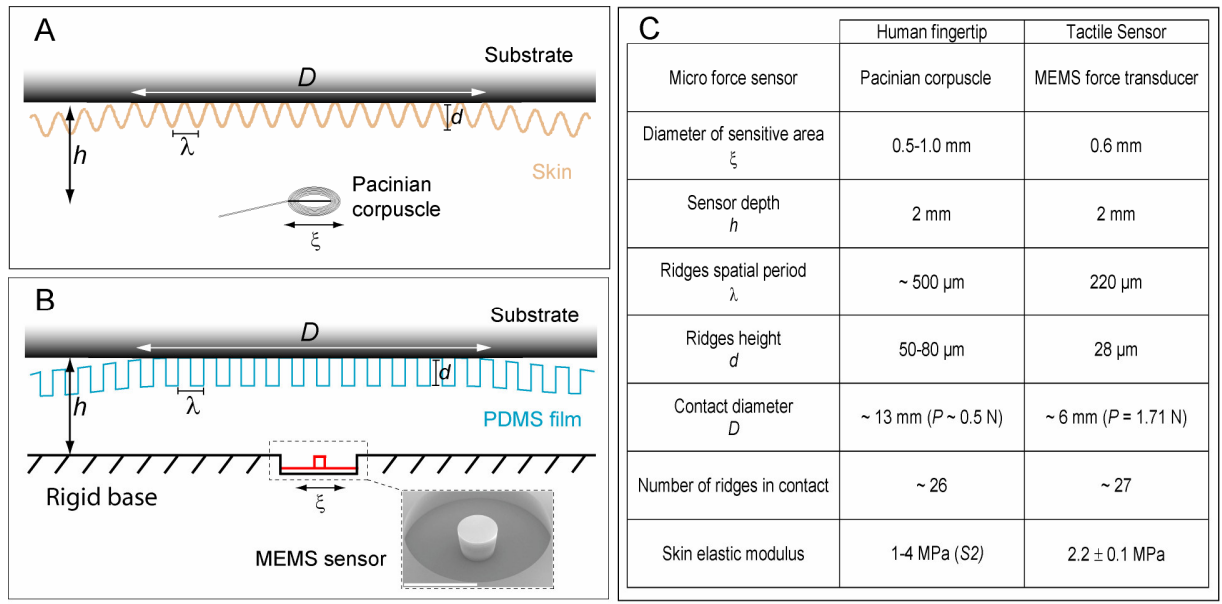

**Fig. S1. Definition and comparison of the geometrical and mechanical characteristics for both biological and biomimetic systems. (A)** Sketch showing a cross-section of the skin surface in a real human fingertip. Since we aim at mimicking the operation of Pacinian corpuscles, only this mechanoreceptor has been represented. **(B)** Sketch showing its equivalent for the tactile sensor. The picture is a detail of the sensitive part of the MEMS sensor. It consists of a joystick-like structure made of a silicon cylindrical post attached to a suspended silicon membrane. Piezoresistive gauges embedded within the membrane allow one to measure its deformations whenever a force is applied on the post. The white bar is 1mm long. **(C)** Table summarizing and comparing the values of key parameters for both systems.

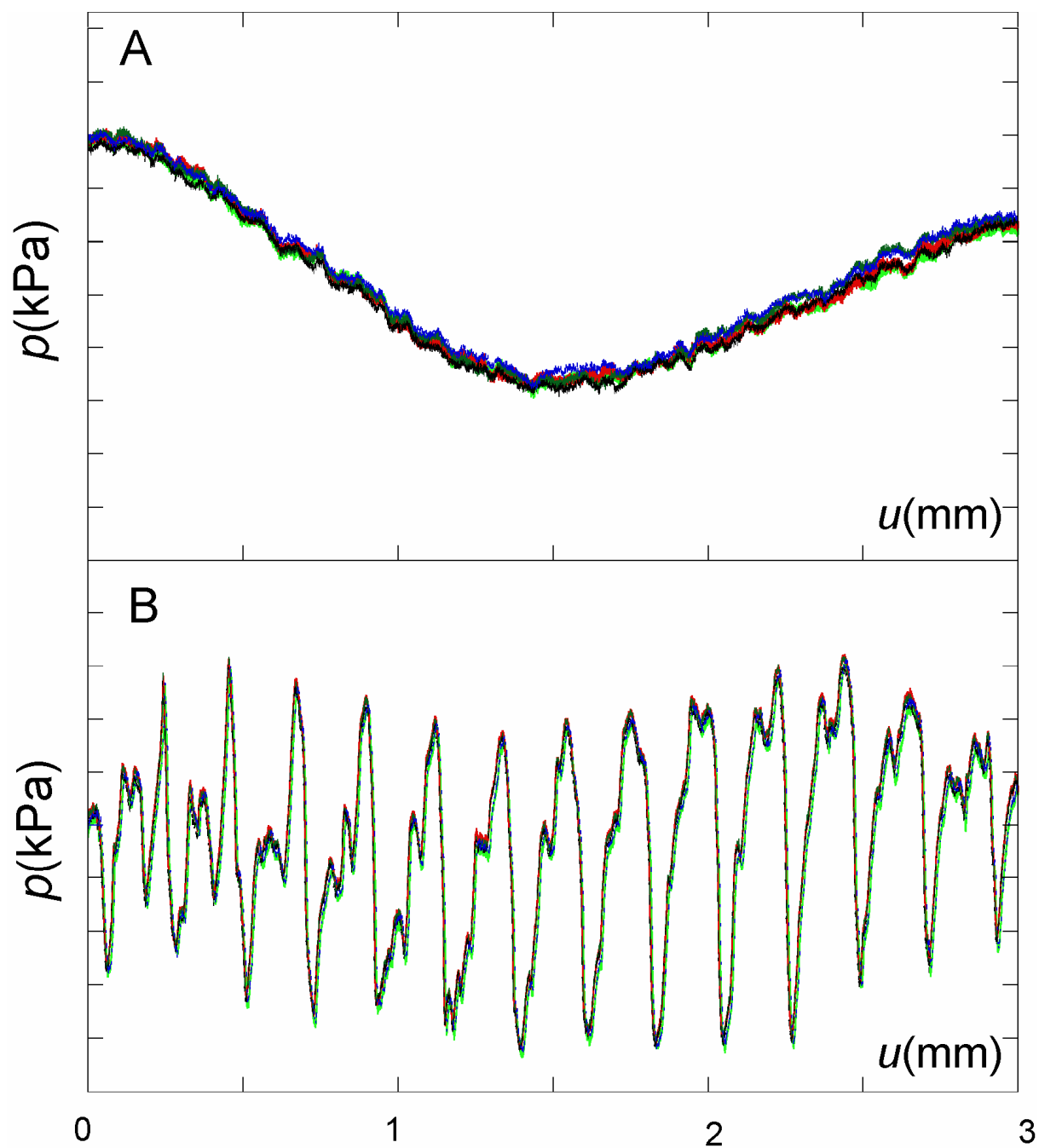

**Fig. S2. Measurements reproducibility.** Five successive recordings of the pressure signal $p$ as a function of the substrate displacement $u$ obtained with **(A)** the smooth and **(B)** the fingerprinted skin sensors. Each interval on the $y$-axis corresponds to a pressure variation of 1kPa.

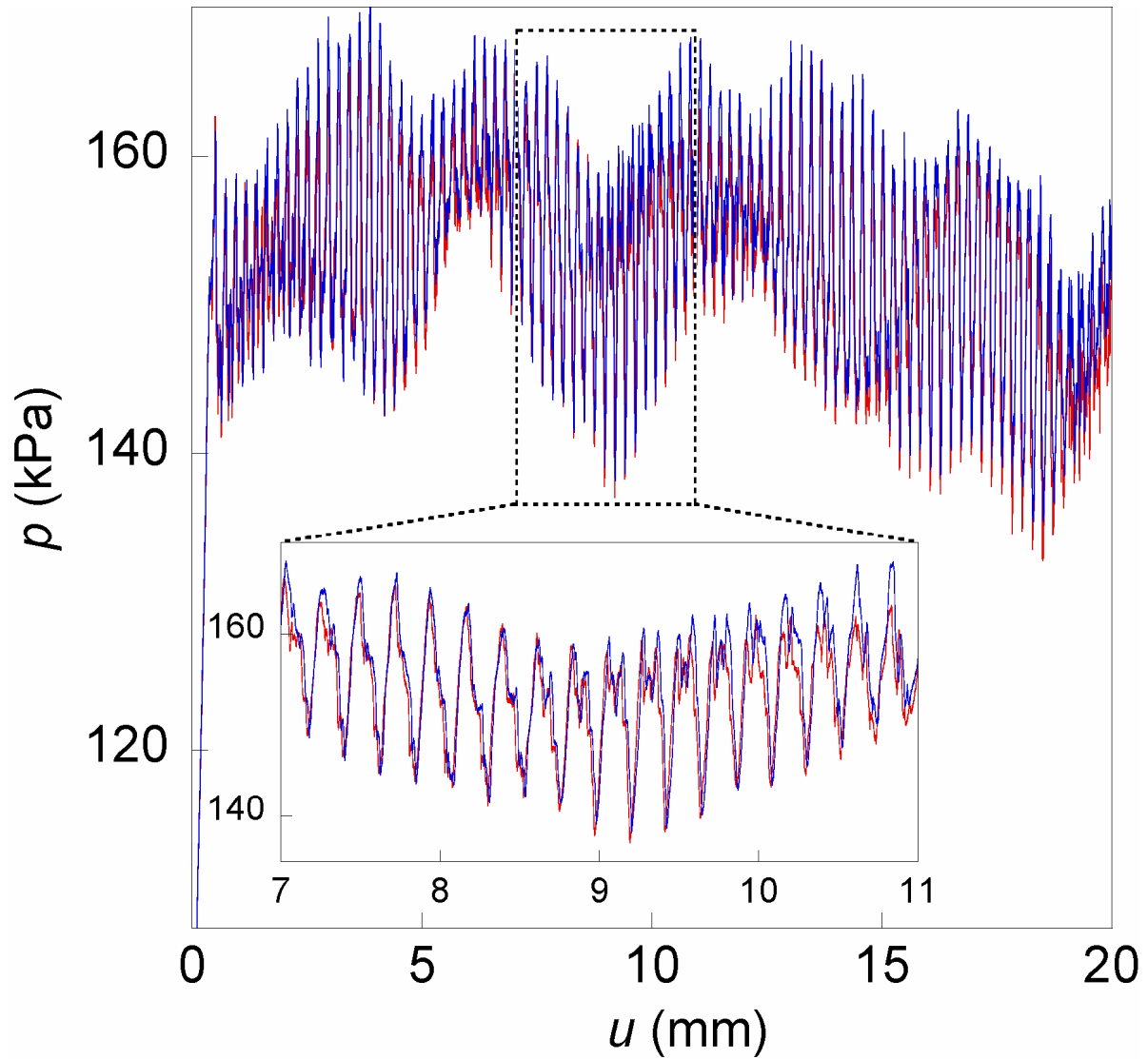

**Fig. S3. Dependence on the scanning velocity.** Two recordings of the pressure signal *p(t)* as a function of the substrate displacement *u = v t* obtained with the fingerprinted tactile sensor, with scanning velocities  *v* = 0.2mm/s (in red) and *v* = 0.4mm/s (in blue). The signals are similar up to fine details as shown in the zoomed in region.

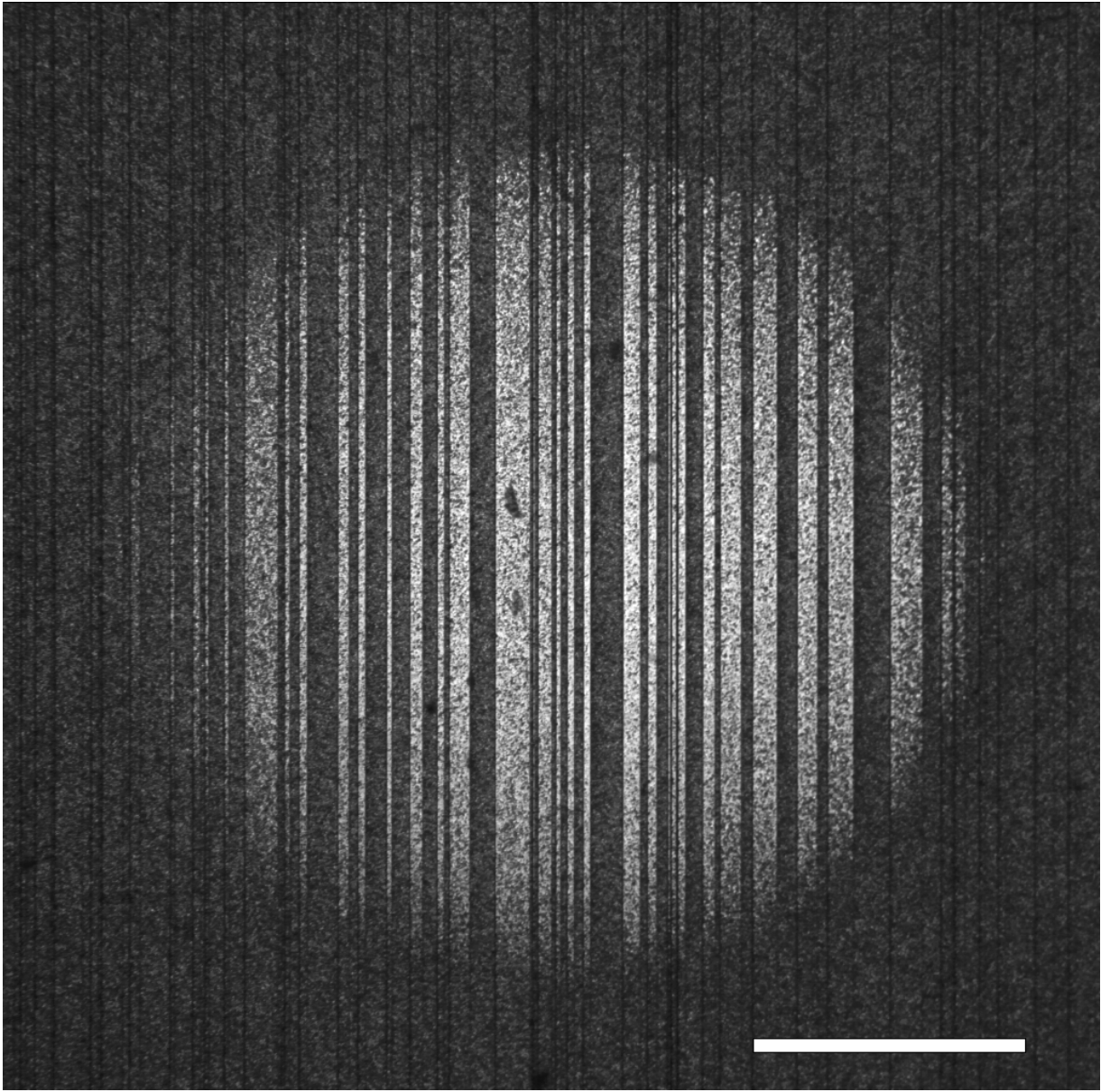

**Fig. S4.** Snapshot of the contact in steady sliding between the smooth membrane and a 1D patterned substrate moving to the right. This image was obtained by imaging in transmission the contact with a white LED and a CCD camera. Regions of actual contact appear bright and are limited to the summits of the substrate grating (no contact occurs over the wells). The white bar is 2mm long.

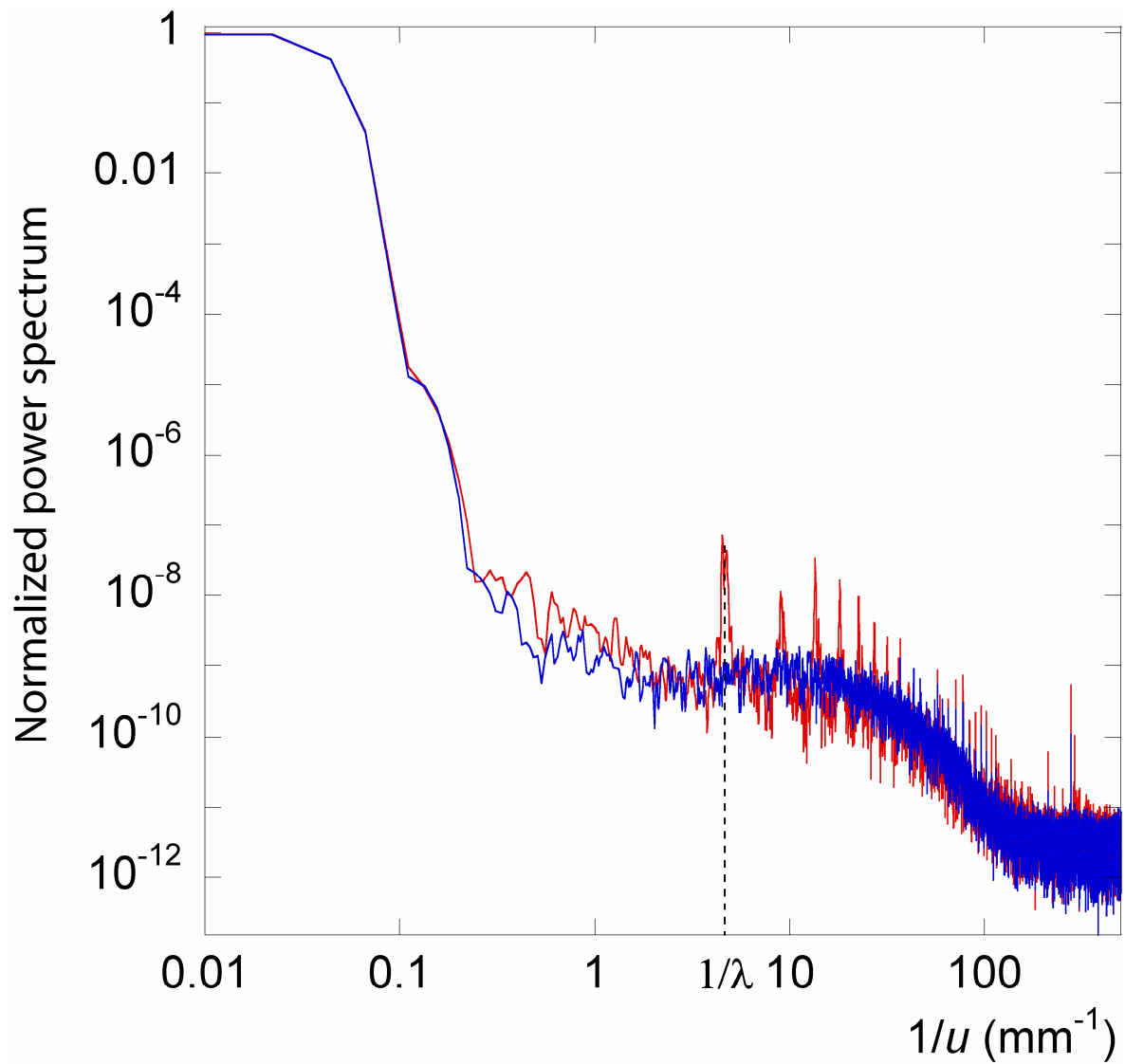

**Fig. S5.** Normalized power spectra of pressure variation signals measured with the fingerprinted (red) and smooth (blue) sensors scanned across finely abraded glass substrates (*u* denotes the substrate displacement and λ the inter-ridge distance). The applied normal load is *P*=1.71N and the scanning velocity is *v*=0.2mm/s. Each spectrum was obtained by averaging over 5 independent scans, each of them 45mm long.

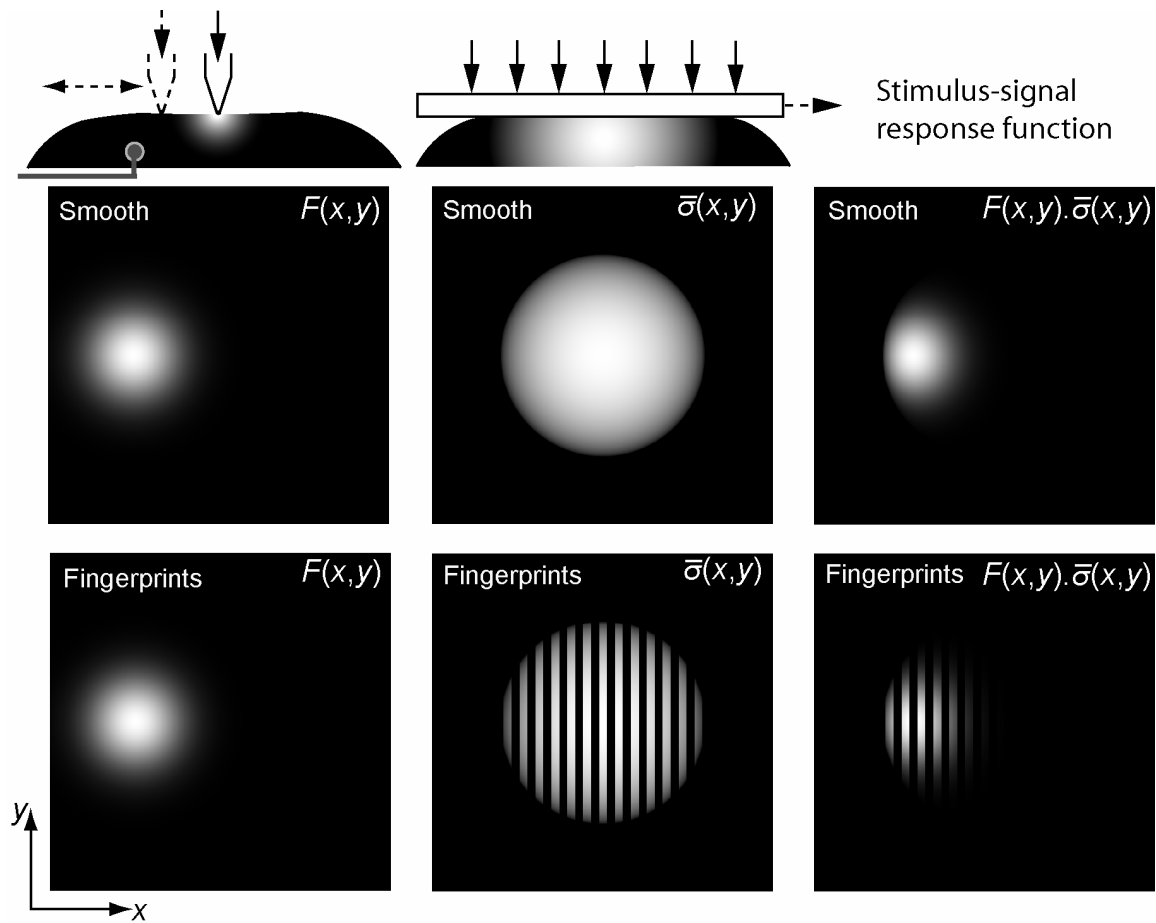

**Fig. S6. Illustration of the response function.** The first column shows a typical receptive field $F(x, y)$ for both smooth and fingerprinted tactile sensors. This function characterizes the intrinsic response of the sensor. It is obtained by measuring the response of the sensor to localized unit forces on the membrane surface. Its typical lateral extension is set by the membrane thickness, but is independent of the membrane fine texture (such as the presence of ridges). $F(x, y)$ is thus identical for both types of membranes. The second column shows the typical reference stress field $\bar{\sigma}$, i.e. the interfacial (time-invariant) stress field produced when rubbing a smooth substrate under a given load and friction coefficient. These parameters set the large-scale shape of $\bar{\sigma}$ but further modulations can be induced by fine textures of the membrane surface. In particular, the presence of epidermal ridges results in a total extinction of $\bar{\sigma}$ along parallel lines as shown. The last column displays the product of $F$ and $\bar{\sigma}$

which defines the linear response function of the sensor (*see* Eq. 2 in the main text). The envelope's extension of $F.\overline{\sigma}$ is mostly controlled by $F$ (although its shape can be distorted by $\overline{\sigma}$). The presence of skin ridges results in short-scale spatial modulations of the response function.

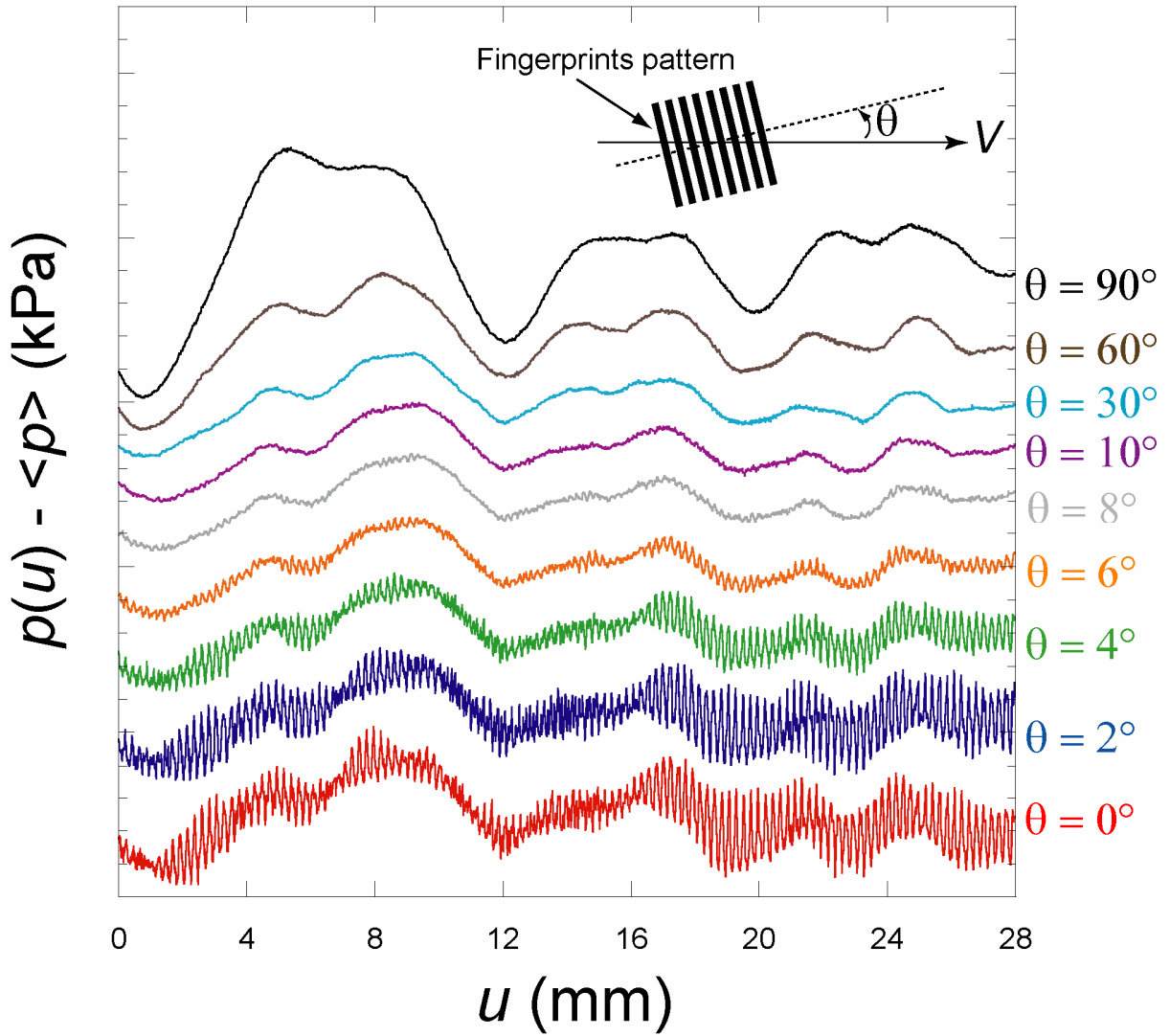

**Fig. S7. Ridges orientation effect.** In this experiment, the tactile sensor was rotated by an angle θ (from 0 to 90°) with respect to the direction of motion and swept across a 1D patterned substrate at $v$ = 0.2 mm/s and with $P$ = 1.71 N. Each interval on the *y*-axis corresponds to a pressure variation of 10kPa. Curves are arbitrarily shifted for visualisation.

# III-References